\documentclass[aps,pra,onecolumn]{revtex4-2}%

\usepackage{graphics}
\usepackage{amsmath}
\usepackage{graphicx}
\usepackage{amsfonts}
\usepackage{amssymb}
\usepackage[normalem]{ulem}
\usepackage{braket}
\usepackage{hyperref}
\usepackage{siunitx}

\begin{document}

\title{Quantum communications feasibility tests over a UK-Ireland 224-km undersea link}

\author{
Ben Amies-King$^{1,\dagger}$,
Karolina P. Schatz$^{1,\dagger}$,
Haofan Duan$^{1,\dagger}$,
Ayan Biswas$^{1}$,
Jack Bailey$^{2}$,
Adrian Felvinti$^{2}$,
Jaimes Winward$^{2}$,
Mike Dixon$^{2}$,
Mariella Minder$^{1,3}$,
Rupesh Kumar$^{1}$,
Sophie Albosh$^{1}$,
Marco Lucamarini$^{1}$\\
}
\affiliation{
$^{1}$School of Physics, Engineering \& Technology and York Centre for Quantum Technologies, Institute for Safe Autonomy, University of York, YO10 5FT York, U.K.\\
$^{2}$euNetworks Fiber UK Limited, 5 Churchill Place, London, E14 5HU, U.K.\\
$^{3}$Department of Electrical Engineering, Computer Engineering and Informatics, Cyprus University of Technology, Limassol, 3036, CY\\
*Correspondence: ben.amies-king@york.ac.uk (B.A-K.), karolina.schatz@york.ac.uk (K.P.S), marco.lucamarini@york.ac.uk (M.L.)\\
$^\dagger$These authors contributed equally to this work.
}

\begin{abstract}
\noindent The future quantum internet will leverage existing communication infrastructures, including deployed optical fibre networks, to enable novel applications that outperform current information technology.
In this scenario, we perform a feasibility study of quantum communications over an industrial 224 km submarine optical fibre link deployed between Southport in the United Kingdom (UK) and Portrane in the Republic of Ireland (IE).
With a characterisation of phase drift, polarisation stability and arrival time of entangled photons, we demonstrate the suitability of the link to enable international UK-IE quantum communications for the first time.
\end{abstract}

\maketitle

\section{Introduction}

\noindent Quantum key distribution (QKD)~\cite{bennett_quantum_2014,gisin_quantum_2002,xu_secure_2020,pirandola_advances_2020} aims to enable confidential communications with security guaranteed by the laws of physics, even in the presence of an eavesdropper with superior computational ability.
Its most common implementation utilises fibre optical channels with C-band light carrying information encoded in polarisation~\cite{BCBB+92} or optical phase~\cite{B92}.
The former degree of freedom is advantageous due to the ease of establishing and maintaining a stable reference between the parties~\cite{agnesi_all-fiber_2019,avesani_stable_2020}, even with passive transmitters~\cite{wang_fully_2023}.
The latter enables beneficial rate-distance scaling via `twin-field' (TF) QKD~\cite{lucamarini_overcoming_2018} and has led to a sequence of QKD demonstrations over record-breaking distances in recent \mbox{years~\cite{minder_experimental_2019,pittaluga_600-km_2021,wang_twin-field_2022,liu_experimental_2023},} overcoming the secret key capacity of a point-to-point lossy channel~\cite{pirandola_fundamental_2017}.
QKD can also be achieved by the distribution of entanglement between distant users, followed by local measurements~\cite{ekert_quantum_1991,bennett_quantum_1992}.
In addition to QKD, entanglement is a fundamental resource for other quantum information protocols, e.g., quantum teleportation~\cite{bennett_teleporting_1993,bouwmeester_experimental_1997,boschi_experimental_1998}.

To date, there have been several QKD field trials on deployed fibre infrastructure worldwide~\cite{PPA+09, SFI+11, DWT+19, JAW+19, Avesani:21, WWL+21, ribezzo_deploying_2023, bersin_development_2023}, although only a limited number demonstrate long-distance international quantum communications~\cite{NBB22}. 
Within these, only one submarine fibre-based communication link~\cite{wengerowsky_entanglement_2019, wengerowsky_passively_2020, ribezzo_quantum_2023} has been reported on, so undersea optical fibres still represent a largely unexplored scenario.
The longest geographical distance achieved in submarine fibre so far, between Italy and Malta, is of around 96~km~\cite{wengerowsky_entanglement_2019, ribezzo_deploying_2023} or 192~km in a loop-back configuration~\cite{wengerowsky_passively_2020}.

In this work, we perform a series of experiments to assess the suitability of a {224}~km submarine fibre link for quantum communication protocols.
The link, denominated `Rockabill', has been deployed by the company euNetworks~\cite{EUN} between their cable landing stations (CLSs) in Southport (UK) and Portrane (IE).
It features low loss ({0.17}~dB/km on average) to enable classical communications between the two endpoints without resorting to intermediate optical repeaters, and low latency to serve time-hungry customers, e.g., from the financial sector.
Hence, the Rockabill link is a natural candidate to connect the UK and IE via quantum communications.
However, this requires experimental verification, to rule out the uncertainty entailed by prior simulations, as reported below.

To this end, we perform a number of tests to characterise the UK--IE channel for quantum communications.
We first carry out a measurement of the relative phase drift between two paired fibres in the channel.
This is the relevant degree of freedom in applications like TF-QKD and time--frequency reference dissemination~\cite{clivati_optical_2020}.
Secondly, we characterise the impact of polarisation drift on the quantum bit error rate (QBER) of a proof-of-concept prepare-and-measure polarisation-based QKD setup operating at the single photon level.
Polarisation drift, inherent to optical fibres, leads to increased QBER and necessitates the periodic realignment of the system.
We explore the timescales over which polarisation is sufficiently stable and attempt to gain insights into the origins of the most prominent noise features.
Finally, we demonstrate the coincidence measurements of photon pairs generated by a compact commercial entangled photon source~\cite{ozoptics_polarization_2023} in the presence of significant channel noise, which is an important step towards the distribution of entanglement over the full length of the optical link, 448~km (2 $\times$ 224~km).

With the combination of these experiments, we report a unique in-the-field feasibility study with wide-ranging implications for a variety of quantum communication protocols, particularly QKD.

\section{Materials and Methods}

\subsection{The Rockabill link}

\begin{figure}[h]
\includegraphics[width=0.95\textwidth]{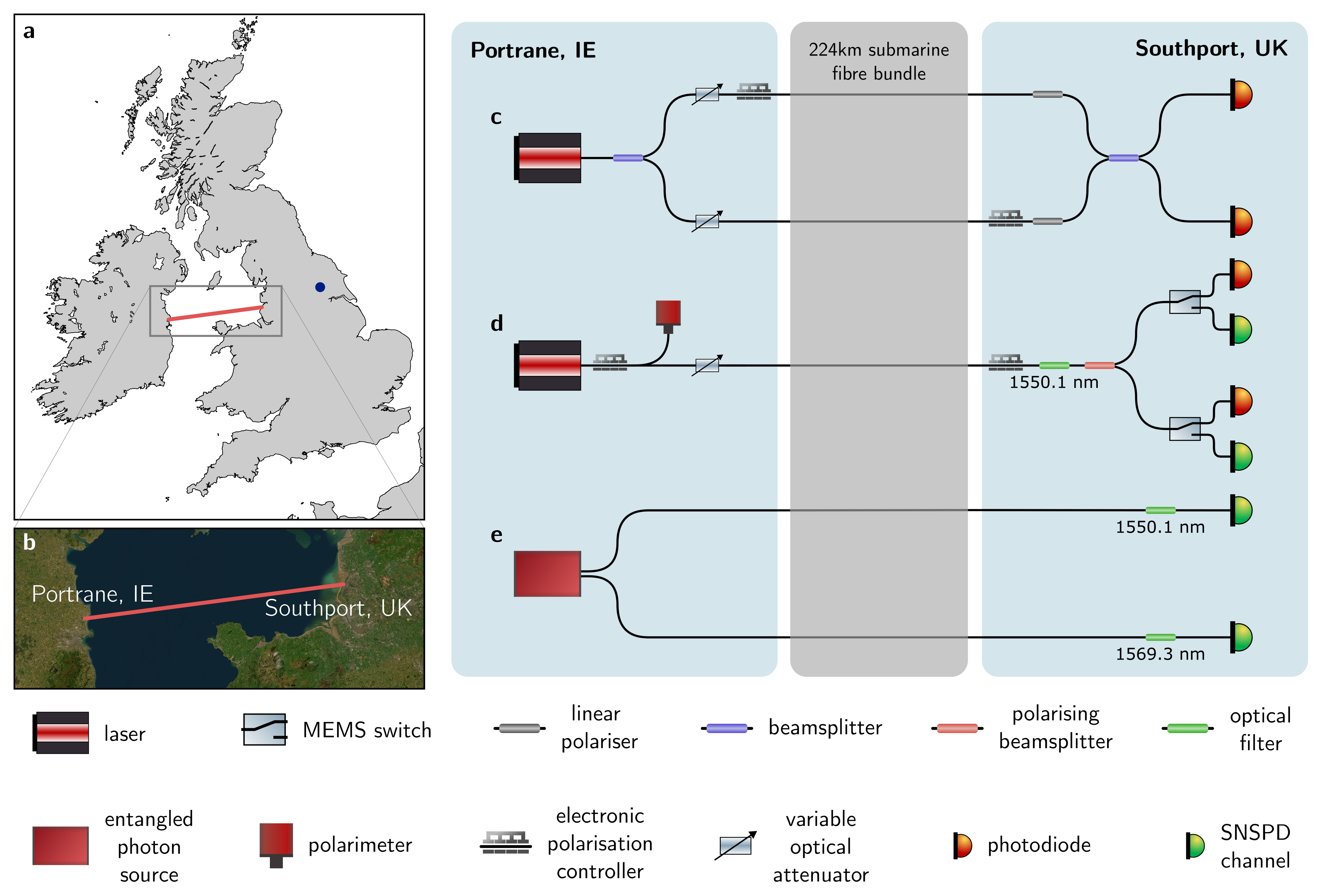}
\caption{Field trial and experimental setup.
(\textbf{a}) Geographic representation and (\textbf{b}) satellite image of the field trial.
The `Rockabill' link deployed by the company euNetworks~\cite{EUN} is drawn as an orange line with endpoints in the cable landing stations (CLSs) of Portrane, IE, and Southport, UK.
The blue dot in the top panel denotes the University of York (UoY).
The whole experimental setup and its subsystems, including the SNSPDs, were moved from the UoY to the two CLSs to perform the feasibility tests.
(\textbf{c}--\textbf{e}) Subsystems of the experimental setup to characterise the optical phase (top), polarisation (middle), and entangled photon pair distribution (bottom) across the UK--IE link.
Map data courtesy of GHSSG, Esri, Maxar, Earthstar Geographics, and the GIS User Community.}
\label{fig:rockabill-map}
\end{figure}

\noindent In this work, we utilise two equivalent fibre pairs (identified as channels 39 and 40; and 59 and 60) in the above-described submarine Rockabill link.
Geographic and diagrammatic representations of the link are provided in Figure~\ref{fig:rockabill-map}.
The fibres under test are routed in a bundle containing several active telecommunication fibres carrying high-intensity signals.
The link has a length of 224 km and a total effective loss per fibre of 38.1--38.7 dB, which equates to a remarkably low average attenuation per kilometre of approximately {0.17}~dB/km.
Despite this low loss coefficient of the fibres, operating at the single photon level across such high channel attenuation is challenging with infrared single-photon avalanche detectors (SPADs), due to the low detection efficiencies and high dark counts, unless special features are applied to the detectors~\cite{FLD+17}.

In contrast, a superconducting nanowire single-photon detector (SNSPD) system offers a much higher detection efficiency and low dark counts at the expense of a larger footprint.
We therefore set up an SNSPD system (ID281 from IDQ~\cite{IDQ_SNSPD}) in Southport and use two detector channels with detection efficiencies of {93.0}\% and {90.5}\%, and dark count rates of less than 70~Hz per channel.

For a number of optimisation procedures required for these experiments, the detection and actuation equipment are on the opposite sides of the fibre link.
To enable these remote processes, we develop and deploy a lightweight asynchronous web server to bridge between Portrane and Southport.
In principle, classical communication could be performed via conventional data signals sent over the fibre link itself.
However, this would introduce noise into the measurements due to scattering effects from bright multiplexed signals, thus making the option of a separate server preferable.

\subsection{Loss and noise characterisation}
\label{subsec:characterisation}

\noindent An extensive attenuation and reflection characterisation of the submarine fibre channel had been provided to us by euNetworks for the channels 39 and 40, stating 38.65 and 38.14~dB as the channel insertion losses, respectively.
We verify the combined loss figure for both channels ourselves by connecting the two fibre outputs in Portrane, effectively creating a loop from Southport back on itself.
In this loop-back configuration, we send {10}~mW of light from a laser into the first fibre, which is looped back into a second fibre and measured on an optical power metre, showing a total loss of 76~dB.
For the other pair of channels 59 and 60, no prior characterisation was available.
We determine the insertion loss by sending {5}~mW of light from Portrane through the fibres and measuring the output at Southport.
This gives us an estimation of 38.2 and {38.6}~dB loss for the channels 59 and 60, respectively, comparable to the attenuation of the other fibre pair.

In addition to losses, the presence of bright active fibres in the same bundle as our experimental fibre leads to significant cross talk in the channels and consequently an unexpectedly high noise floor.
We perform a number of tests to characterise the source of noise and its influence on the quantum signals.
During these tests, the entrance to the channels in Portrane are terminated, and different instruments are connected in the darkened CLS in Southport to investigate the light arriving there.

Firstly, we record a {12}~h-long overnight measurement of the single photon count rate on a single channel using a SPAD.
The measurement includes sunset as well as sunrise and shows a remarkable long-term stability of the noise amplitude (standard deviation $<$ 0.5\%).
Next, we perform a coarse and a fine sweep of the noise wavelength spectrum using a tuneable optical wavelength filter directly before the SNSPD, at bandwidths of {125} and {100}~GHz.
The results of the coarse sweep for two selected channels can be seen in the inset to Figure~\ref{fig:EPS_results}.
The spectrum is not indicative of a typical scattering process or thermal emission; instead, it resembles a convolution of several peaks of varying amplitude and width.
This test also serves to exclude the possibility that stray light enters the exposed fibre within the CLS itself, since disconnecting the fibre channel from the measurement setup, while leaving all the other fibres in place, removes all counts on the SNSPD apart from the typical {70}~Hz detector dark counts (see grey line in Figure~\ref{fig:EPS_results} inset).
Furthermore, a quick investigation using a manual polarisation controller followed by a linear polariser shows that the noise in its entirety does not have a preferred polarisation orientation, whereas individual peaks isolated with the tuneable optical filter can be found to be at least partially~polarised.

Combining the above results leads us to the conclusion that the noise is indeed caused by stray photons from other active fibres entering our channels under test, most likely from the back-reflection of bright signals leaking into the tested channels close to the detection setup in Southport.
We find such a background to be present in all channels arriving at the CLS, but its magnitude varies significantly from channel to channel, most likely depending on physical proximity to the most active fibres.
For the two fibre pairs used in this work, we measure the count rates ranging from {120}~kHz to {410}~kHz when permitting the entire spectrum to the SNSPD.
Introducing optical filters in the order of a {1}~nm bandwidth significantly mitigates this noise and makes our quantum measurements possible.
Effective dark counts for the single-photon experiments reported here range from {90}~Hz to as high as {2.2}~kHz, depending on the chosen central wavelength of the filters utilised.

\subsection{Optical phase characterisation}
\label{subsec:opt_ph_setup}

\noindent For the first experiment, a {1}~kHz linewidth laser at {1550}~nm is coupled into the fibre pair via a 50:50 beamsplitter in Portrane, as shown in Figure~\ref{fig:rockabill-map}{c}.
One output of the beamsplitter (arm 1) is connected to the channel via an electronic polarisation controller (EPC1) and both outputs have an electrical variable optical attenuator (EVOA) used to enable polarisation alignment.
In Southport, the channel that is not connected via EPC1 in Portrane (arm 2) enters an identical EPC (EPC2).
Both arms then pass through linear polarisers before interfering on a beamsplitter.
The beamsplitter outputs are measured on high-sensitivity photodiodes at a sampling rate of 2~MS/s.

Before running an acquisition, polarisation is aligned to optimise the interference visibility.
A gradient descent algorithm maximises the power on the photodiodes in the absence of interference, after fully attenuating the light from the complementary arm.
This process is performed individually and sequentially per channel using the EPCs.

A similar process is used to obtain the detector noise shown in Figure~\ref{fig:phase}b; with one arm fully attenuated and polarisation aligned, an acquisition is run in the absence of any interference and thus without phase noise from the channel.

\subsection{Polarisation characterisation}
\label{subsec:polarisation_setup}

\noindent  In order to characterise the impact of channel polarisation drift on the QBER, the setup is modified to only include a single fibre in the undersea channel.
Known polarisation states are prepared at the transmitter using EPC1 and a polarimeter, connected to the now vacant end of the 50:50 beamsplitter.
The desired polarisation state is attained through a gradient descent algorithm.
Polarisation alignment between Southport (receiver) and Portrane (transmitter) is performed by preparing a horizontal state at Portrane and then maximising the extinction ratio of the optical power at the two outputs of the polarising beamsplitter (PBS) in Southport using EPC2.
This process is performed with classical light with an optical power in the order of hundreds of nanowatts.
Once alignment is complete, the desired state for each measurement can be prepared in Portrane based on the polarimeter reading.

Both state preparation and alignment introduce experimental errors due to the relatively high tolerance in the alignment algorithms, which is made necessary by the low voltage granularity that the EPCs can be addressed with.
In the case of state preparation, the alignment algorithm terminates once the prepared state is less than $3^{\circ}$ (0.05 rad) away from the target state on the Poincar\'e sphere as measured by the polarimeter.
For the detector alignment, the extinction ratio between the two outputs of the PBS is {25}~dB, which would ideally generate a base QBER of $\sim$0.3\%.
However, based on the observed performance of the EPCs, achieving such an alignment requires running the optimisation for a long time.
We limit the execution time of the gradient descent algorithm to less than a minute, which returns an average base QBER of $\sim$1.6\%, still well within the tolerance of QKD.

Firstly, we characterise the long-term polarisation drift over the course of nearly {14}~h using classical photodetectors.

In order to determine the channel's QBER for the four polarisation states used in the BB84 protocol, we attenuate the power launched into the channel to the single photon level after alignment.
In this regime, the continuous wave laser light is set to {128}~pW, which is equivalent in power to sending single photon pulses of exactly $\mu=1$ at a repetition rate of {1}~GHz.
Our laser is attenuated via a series of electrical variable optical attenuators (EVOAs) to achieve the desired launch power for $\mu=1$, whilst the output power of the laser itself is regulated via an internal feedback system.
We obtain the true power launched into the channel by measuring the attenuation of each EVOA and the optical power directly before the final EVOA.
Additional measurements at a fraction of single photon power are also performed to explore amplitude-varied state generation in relation to the decoy states technique~\cite{Hwa03,Wan05,LMC05}.

After the desired optical power is achieved, polarisation-maintaining micro-electromechanical system (MEMS) optical switches are used to redirect the light from the classical detectors to the SNSPD channels.
The two SNSPD channels are connected to a time tagger, which records the single photon count rates for {300}~s per measurement.

\subsection{Distribution of photon pairs characterisation}
\label{subsec:entanglement_setup}

\noindent In this feasibility test, we explore the channel's capability to distribute photon pairs produced by a compact source of entangled photon pairs (EPS, OZ Optics `Ruby').
For this, we adopt a subsystem of the setup where the laser of the previous test is replaced by the EPS with outputs connected to two parallel, although distinct, fibres of the UK--IE channel.

Photon pairs are launched from the Portrane CLS into two {224}~km fibres towards the Southport CLS, where they are detected with the SNSPDs connected to a time tagger.
From the time tags, we reconstruct the photons' arrival times and identify single and coincidence~counts.

The EPS produces polarisation-entangled photon pairs, created via spontaneous parametric downconversion (SPDC) in an auto-balanced displacement interferometer~\cite{HJ19}.
The interferometer consists of two {20}~mm\-long type-0 phase-matched periodically poled lithium niobate (PPLN) crystals, pumped by a {780}~nm continuous wave laser at a maximum power of 4.4 mW.
Signal and idler single photons are created in the maximally entangled Bell state $\ket{\Phi^+} = (\ket{HH} + \ket{VV})/\sqrt{2}$ with an entanglement fidelity $>$98\%.
The source's wavelength spectrum exhibits a full width at half-maximum of approximately {60}~nm, centred around {1560}~nm.
Energy conservation requirements dictate that the entangled photons are correlated in wavelength and equidistant from {1560}~nm, in this case with the idler and the signal at higher and lower wavelengths, respectively.

In Southport, optical filters centred at {1550.1}~nm and {1569.3}~nm with equal pass bands of $\Delta$~=~{1.39}~nm are used to isolate entangled pairs.
This both reduces the time-of-flight-induced broadening of our measured photon arrival times due to chromatic dispersion and combats the cross talk in the channel.
Note that the idler side experiences roughly {5}~dB of additional attenuation compared to the signal channel due to a high-insertion-loss optical filter, which is reflected in the detected count rate.
We take advantage of the fact that the noise spectrum has a minimum of approximately {1570}~nm for the idler detection, which coincidentally puts the  signal wavelength at the conventional telecom wavelength of {1550}~nm.
Even so, in this experiment the link noise is comparable to the signal amplitude, as can be seen in Section~\ref{subsec:characterisation}.

Since the SNSPD efficiency is polarisation-dependent, we also perform a polarisation alignment procedure to maximise the probability of detecting paired photons.
This procedure consists of blocking one half of the source interferometer to produce the non-entangled two-photon state $\ket{\Psi} = \ket{HH}$ and using manual polarisation controllers to align this state to the SNSPD’s maximum efficiency state.
Due to the very low absolute count rates arriving in Southport, this optimisation cannot be reliably performed, which results in small discrepancies between the two measurements taken here.

\section{Results}

\subsection{Optical phase characterisation}
\label{subsec:opt_ph_meas}

\begin{figure}[h]
\includegraphics[width=0.95\textwidth]{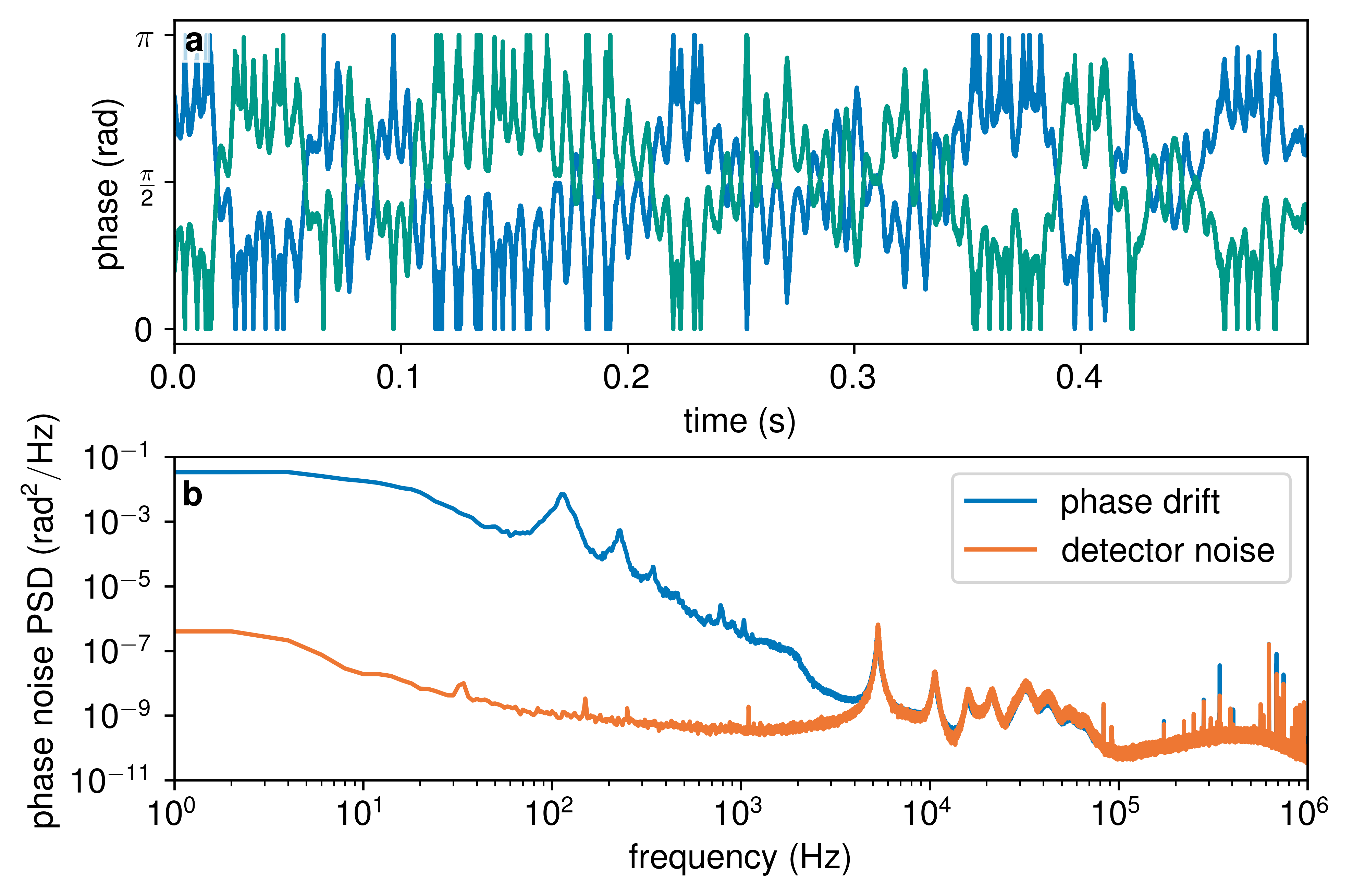}
\caption{{Relative} 
 phase drift of the fibre pair 39 and 40 of the Rockabill link:
(\textbf{a}) \SI{0.5}{\second} of the phase drift of the channel; and
(\textbf{b}) power spectral density (PSD) of the phase noise of the channel (blue), with detector noise (orange) plotted for comparison.\label{fig:phase}}
\end{figure}

\noindent The results of the relative phase drift measurement over the pair of fibres in the Rockabill link are shown in Figure~\ref{fig:phase}.  
Figure~\ref{fig:phase}a 
 shows a {0.5}~s excerpt of the wrapped phase drift of the fibre pair, with clear complementarity between the two lines, which correspond to the two detectors at the outputs of the interfering beamsplitter.
This is representative of the first-order interference taking place over the whole channel length. 
The phase power spectral density (PSD) of the phase drift over {96}~s is plotted in Figure~\ref{fig:phase}b, along with the corresponding phase PSD of the detector in the absence of channel phase noise.

Both subplots show a consistent periodic phase fluctuation at a frequency of {100}~Hz, with a magnitude of a little under $\frac{\pi}{2}$.
Aside from this peak, and a smaller peak at around {125}~Hz, the phase noise spectrum exhibits a roll-off between {10}~Hz and {1}~kHz, above which frequency it quickly converges to the detector noise.
This is around two orders of magnitude lower in frequency than the typical upper limit of the absolute phase drift of a fibre~\cite{li_twin-field_2023}, a phenomenon that agrees with previously observed data and underlines the attractiveness of submarine channels for QKD purposes.

These results are promising, as phase drifts at such low frequencies can easily be compensated through various phase stabilisation techniques implemented in either hardware~\cite{pittaluga_600-km_2021, clivati_coherent_2022} or software~\cite{li_twin-field_2023}.

\subsection{Polarisation characterisation}
\label{subsec:polarisation_meas}

\begin{figure}[h]
\includegraphics[width=\textwidth]{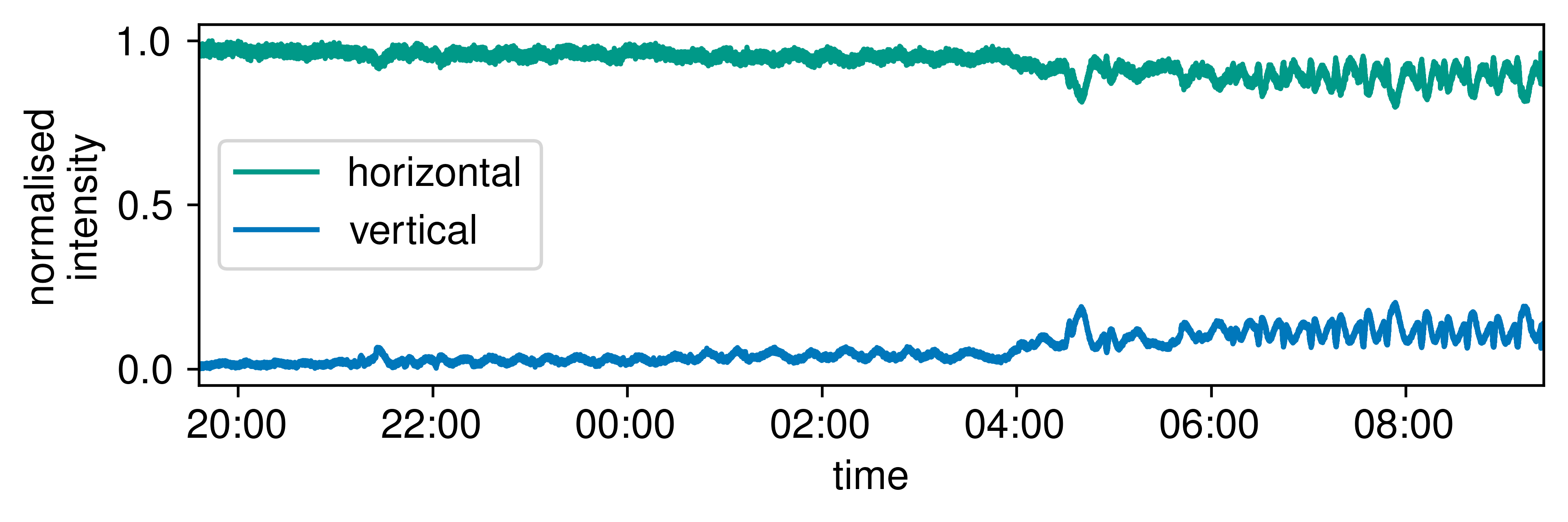}
\caption{The long-term polarisation drift of the channel over \SI{14}{\hour} overnight.}
\label{fig:long_term_polarisation}
\end{figure}

The polarisation drift of the channel is shown in Figure~\ref{fig:long_term_polarisation}.

The measurement was performed over the course of {14}~h, between 19:35 on {5 July 2023} 
 to 09:35 on 6 July 2023.
From the figure, it is apparent that the data remain stable for about {9~h, until approximately 04:30.
These data allow us to approximately extract the interference visibility and hence the QBER, with the latter being between {3.0}\% and {6.7}\%, which falls well within the tolerance limits of QKD.

At around 04:30, the polarisation starts to drift markedly, possibly due to a singular disruption in one of the CLSs (for example, a sudden change in the air conditioning).
It is important to note that the increase in small, almost periodic drifts in polarisation after 4:30 on a minute scale should not be traced back to increased external disturbances, but instead represent the more obvious fluctuation in a non-rectilinear polarisation state.
The later polarisation drift data return an approximate QBER above {10}\% and do not change significantly for the remainder of the measurement.
While this QBER is too large for the transmission of a quantum key, it only occurs after approximately 9 h of unattended operations. 
It would be straightforward to bring the polarisation back to its initial value by repeating the alignment procedure.
These data show a remarkable polarisation stability of the channel, which is not uncommon in undersea fibres.
The polarisation drift is also notably found to be independent of the time of day, due to the CLS's excellent optical isolation and temperature regulation.
Neither sunrise nor sunset at around 05:00 and 22:00, respectively, cause any obvious changes in the polarisation nor in its drifting rate.
Our measurement shows that the alignment procedure would only need to be repeated on an hourly scale, which greatly simplifies the implementation of future quantum communication protocols on this channel.

As explained in Section~\ref{subsec:polarisation_setup}, after the first characterisation of the polarisation over long timescales using bright light and sensitive power meters, we perform measurements over {300}~s with the intensity of light reduced to the single photon level and received by single-photon detectors.
This provides us with a more accurate QBER estimation, which can be used to assess the suitability of the Rockabill link for QKD.

The results are reported in Figure~\ref{fig:snspd_qber}, showing the measured count rates and QBER. 
In Figure~\ref{fig:snspd_qber}{a}--{h}, 
 referring to the preparation and measurement of the polarisation basis $H/V$, the photon flux decreases from left to right.
Correspondingly, the count rates decrease and QBER increases on average, as expected.  
The average QBER when $\mu = 1$ is $2.98 \pm 0.54 \%$ for a $\ket{H}$ preparation $1.55 \pm 0.15 \%$ for a $\ket{V}$ preparation.
Even with the lowest launch power, $\mu = 0.4$, the average QBER is $4.26 \pm 0.67 \%$ and $8.49 \pm 0.96 \%$ for the horizontal and vertical polarisation states, respectively, which are within the expectations for a 224 km optical fibre.
In Figure~\ref{fig:snspd_qber}{f}, there is a more pronounced fluctuation on the scale of minutes, which we attribute to the free evolution of non-perfectly-aligned polarisation states at the ends of the optical link.

To check the polarisation alignment between the four states of the BB84, we also perform a measurement in the $H/V$ basis after preparing the initial states in the $D/A$ basis.
The results are shown in Figure~\ref{fig:snspd_qber} i, j for a photon flux $\mu=1$.
In the ideal case, the QBER would be aligned around the 50\% value.
Our measured QBER is not far from this condition, especially for the preparation of the $\ket{H}$ state (Figure~\ref{fig:snspd_qber}{i}).
In these diagrams, the polarisation fluctuations are more pronounced due to the complementarity of the bases, which entails a linear dependence of the counts on the misalignment angle.

\begin{figure}[h]
\includegraphics[width=\textwidth]{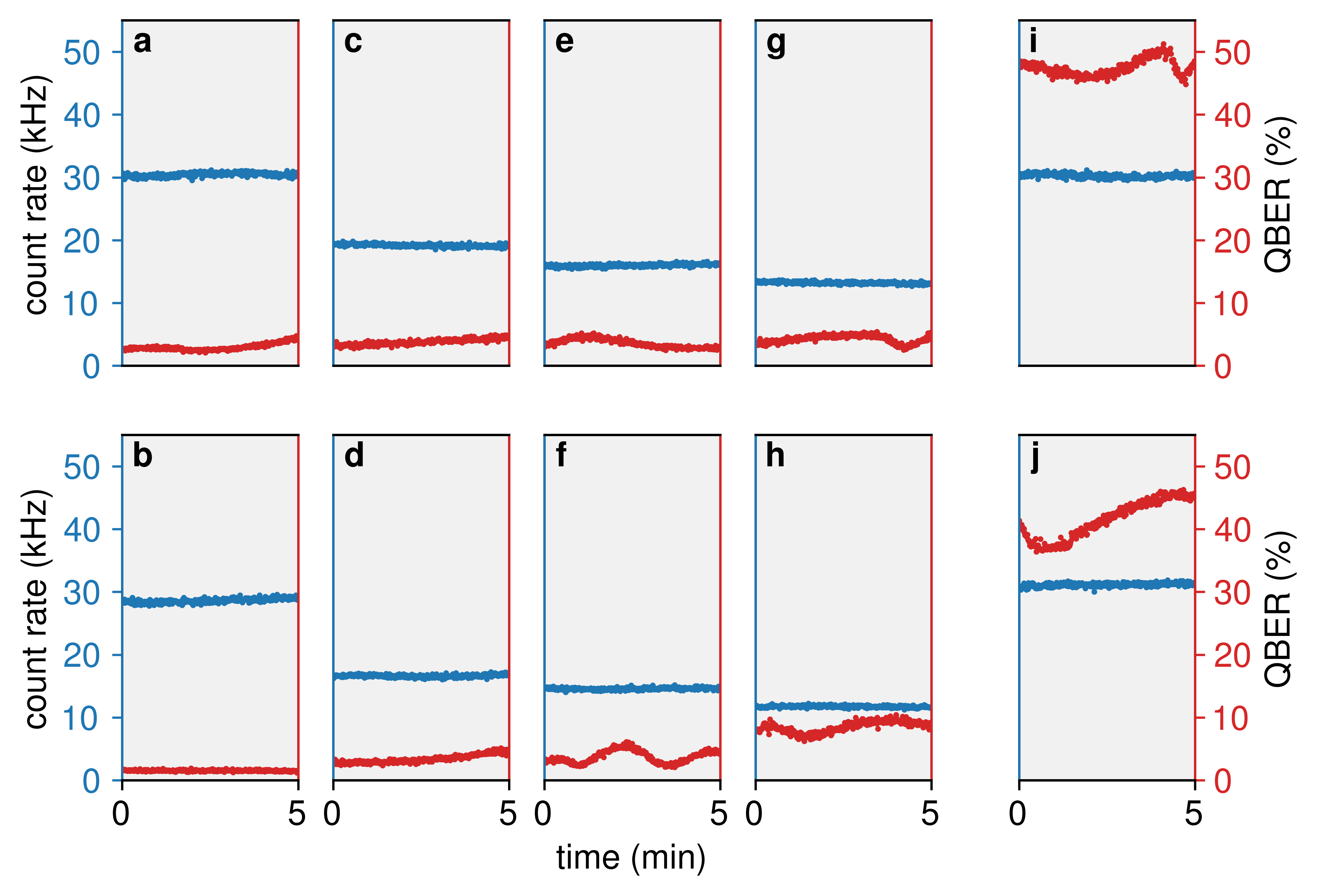}
\caption{Total count rates (blue) and QBER (red) for different quantum states prepared in Portrane, as measured in Southport. 
Plots (\textbf{a}, \textbf{c}, \textbf{e}, \textbf{g}) correspond to a preparation of the state $\ket{H}$ measured in the rectilinear basis $H/V$; plots (\textbf{b}, \textbf{d}, \textbf{f}, \textbf{h}) are similar, but for the state $\ket{V}$. 
Plots (\textbf{i}, \textbf{j}) correspond to the states $D$ and $A$ prepared, respectively, measured in the rectilinear basis $H/V$.
The first and last columns were generated with photon flux $\mu=1$, whereas columns 2--4 depict the states at $\mu$ equal to 0.6, 0.5, and 0.4, respectively.
\label{fig:snspd_qber}}
\end{figure}

The values reported in Figure~\ref{fig:snspd_qber} for a varying photon flux can be used to experimentally infer the asymptotic secret key rate (SKR) of a QKD system clocked at {1}~GHz and running the decoy-state BB84 protocol over the Rockabill link.
Any combination of two single photon intensities will result in a different SKR using the decoy-state method, which presents an optimisation problem.
After plugging the experimentally obtained values of the photon flux, count rates, and QBER in the decoy-state equations~\cite{ma_practical_2005}, we obtain a maximum SKR when signal and decoy states photon fluxes are equal to 0.6 and 0.5, respectively.
This corresponds to an SKR of $2.93\times10^{-6}$~bit/clock and is shown with an orange diamond in Figure~\ref{fig:sim}.
It is expected that this SKR can be considerably improved by further optimising the initial values of the photon flux.
The optimisation should also lead to a positive SKR in the more challenging finite-size scenario, which we leave for future investigations.

Using $\mu=0.6$ and the parameters of the Rockabill link, we also report a simulation of the SKR for a variable distance between the end points Alice and Bob.
The result is given by the blue line in Figure~\ref{fig:sim}.
On one hand, it shows that the simulation matches our experimental observation; on the other hand, it also shows how challenging the experiment is and underlines the necessity to conduct an experimental test.
The length of the Rockabill link falls in the region where the SKR starts to roll over and goes to zero with a steep slope, leading to a zero SKR at {233}~km.
This is only {9}~km away from our experimental point.
To put this figure into context, it would suffice to add one additional beam splitter in Bob's setup to add 3~dB of loss to the channel and result in a SKR of zero.
\begin{figure}[h]
\includegraphics[width=0.75\textwidth]{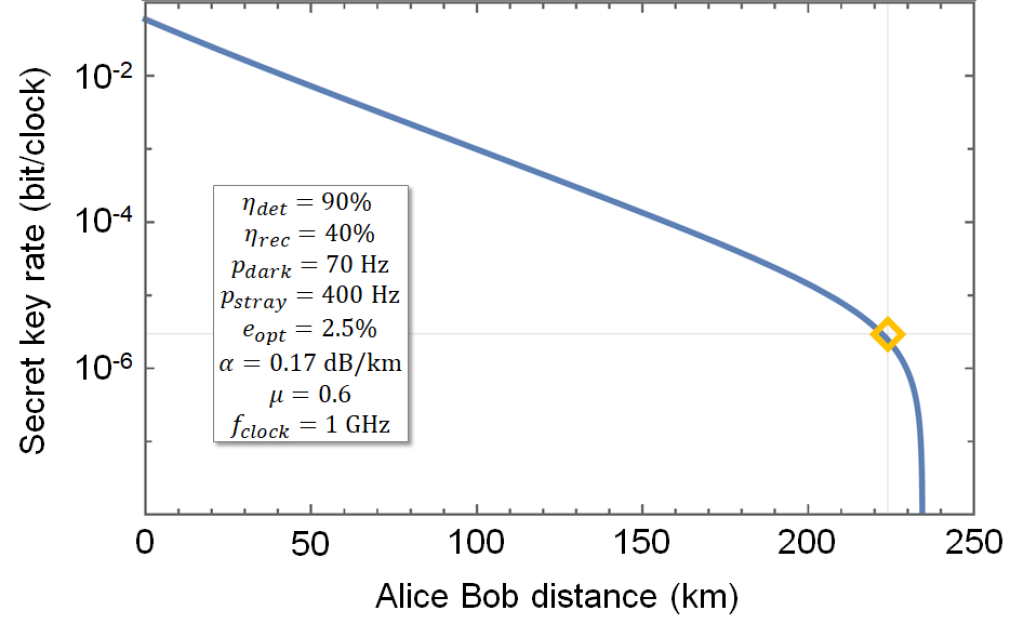}
\caption{{Simulation} 
 of the secret key rate (SKR, blue solid line) with the parameters of the UK--IE field trial (inset) and our experimental point (orange empty diamond). 
The distance of the field trial between Alice, in the Portrane CLS, and Bob, in the Southport CLS, is 224~km, which is only 9~km shorter than the maximum distance providing a positive SKR (233~km).  
Parameters---$\eta_{det}$: detection efficiency (SNSPD); $\eta_{rec}$: efficiency of the receiver; $p_{dark}$: dark counts; $p_{stray}$: background photons from the channel; $e_{opt}$: optical noise; $\mu$: mean photon number; $f_{clock}$: clock rate of the simulated system. 
}

\label{fig:sim}
\end{figure}

\subsection{Photon pair distribution characterisation}
\label{subsec:ent_meas}

\noindent Using the entanglement distribution setup, described in Section~\ref{subsec:entanglement_setup} and Figure \ref{fig:rockabill-map}{e}, we record two sets of data which accumulate single-photon detection events over {12}~h each, time tagged to a resolution of {1}~ps with a combined instrument jitter of the order of {50}~ps.
After traversing the submarine channels, the single count rates of each individual channel are only a few hundred Hz above the background, resulting in a signal-to-noise ratio below one for the noisier {1550.1}~nm channel.
Results are given for both sets, but only the better aligned measurement is discussed in detail in the following.

In the better aligned measurement, the signal channel at {1550.1}~nm records a single count rate of {2703}~Hz over a background rate of {2171}~Hz, whereas the much lower noise idler channel at {1569.3}~nm produces a total count rate of {258}~Hz over  the {88}~Hz background.

Figure~\ref{fig:EPS_results} illustrates the difference in photon arrival times on both channels compared to a common reference clock for the better aligned dataset.
A fixed offset of {113}~ns is present due to minor channel length differences (mainly introduced by our instruments), but the peak corresponding to the simultaneously produced pair of photons is clearly visible.
Based on the low pump power and high entanglement fidelity of our source, this peak is expected to be made up of a vast majority of entangled photons and contributions from multi-photon pairs are expected to be negligibly low.

Photon pairs can be identified in a coincidence window of {12}~ns, which matches the expected broadening through chromatic dispersion after {224}~km.
The observed number of accidental photon coincidences corresponds to the statistically expected value based on the individual count rates.
The coincidences rise above this baseline with a coincidence to  an accidental ratio of 1.92 in the better aligned measurement and 1.80 in the other.
The presence of a clear peak in Figure~\ref{fig:EPS_results} is promising and paves the way for future stronger demonstrations of the entanglement distribution, ultimately prevented in this experiment by the excessive noise described above.
Solutions to further filter the noise and detect entanglement on the same optical link are currently being vetted.
\begin{figure}[h]
\includegraphics[width=0.9\textwidth]{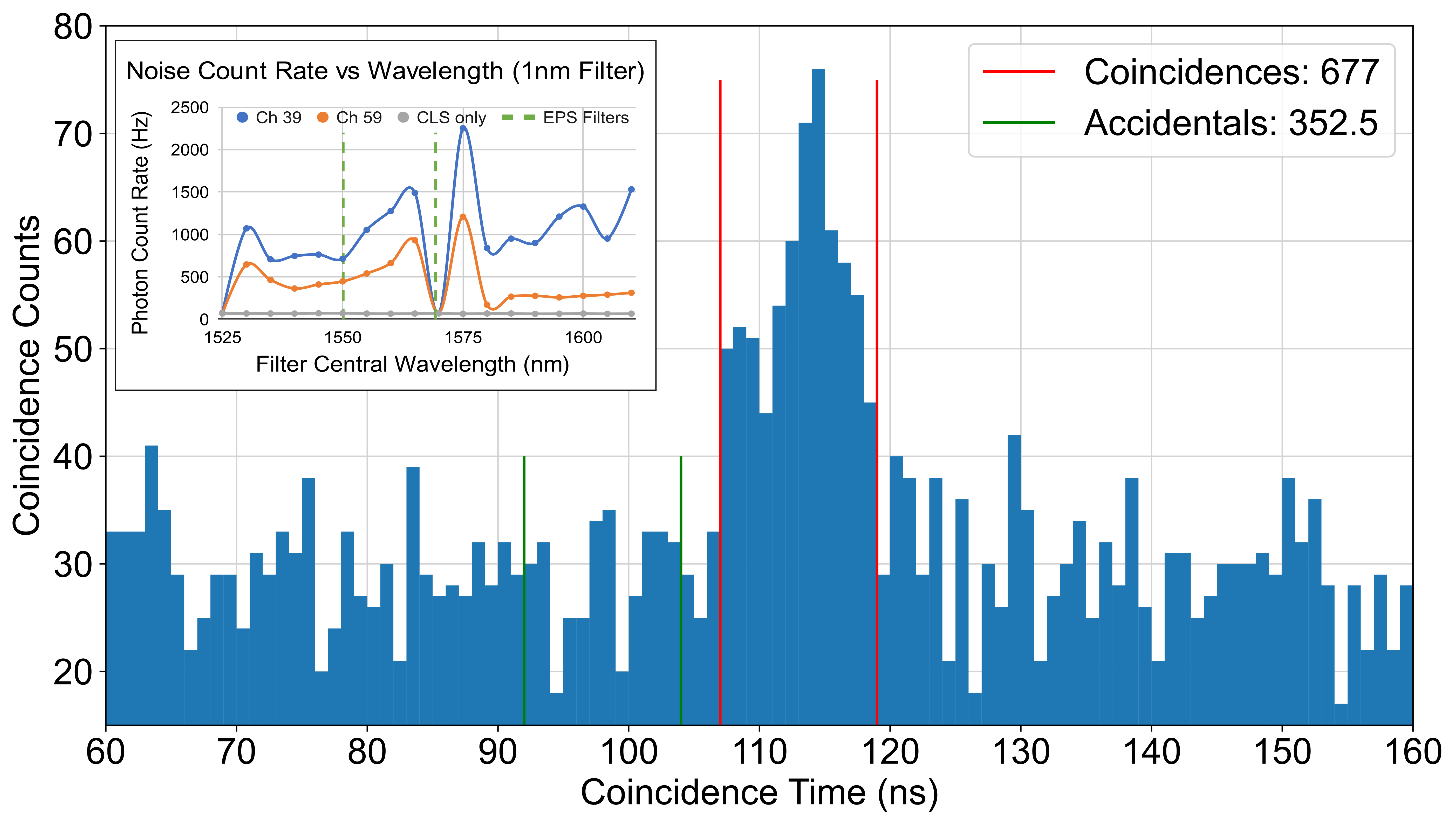}
\caption{Histogram of the time delays between consecutive photon detections at the end of the two submarine channels. Pair photons from the EPS arrive with a fixed relative delay of \SI{113}{\nano\second} at a signal-to-noise ratio of 1.92, computed over the indicated red \SI{12}{\nano\second} coincidence window. The green lines indicates an exemplary coincidence window including only accidentals. The inlay shows the typical spectrum of the stray photon noise present in the channel, with the dashed lines indicating the wavelengths chosen for the EPS measurement.}
\label{fig:EPS_results}
\end{figure}

\section{Discussion}

\noindent We planned, designed, and implemented an experimental campaign aimed at characterising the suitability for quantum communications of the `Rockabill' submarine optical link, deployed by the company euNetworks between Portrane, IE, and Southport, UK.
During the tests, we measured a continuous phase drift, reported in Figure~\ref{fig:phase}, which is significantly below the expected values for a typical optical fibre and in line with previous observations of submarine fibre links.
The reported phase variations are within the range of common phase stabilisation techniques, promising the successful implementation of phase-based QKD~protocols.

The polarisation behaviour of the current link (Figure~\ref{fig:long_term_polarisation}) is also promising for the implementation of a polarisation-encoded BB84 QKD protocol.
The polarisation remains stable over the course of about {90}~minutes.
After this time, it is necessary to re-align the system to assure continuous operation, which we consider straightforward given the long lapse of time.
Over this period of time, the QBER remains below $4.26\%$ for the horizontal state $\ket{H}$ and below $8.49\%$ for the vertical state $\ket{V}$, both of which are sufficiently low values for the implementation of this QKD protocol.
This makes it possible to extract a positive SKR from the data using, in particular, the two $\mu$ values 0.6 (signal state) and 0.5 (decoy state).
It is worth mentioning that other values of the photon flux, e.g., 1 for signals and 0.5 for decoys, would not lead to a positive SKR, showing a non-trivial role played by the selected parameters in the obtained results.

Lastly, we detected the presence of photons emitted by a source of polarisation-entangled photons after a combined propagation distance of {448}~km covered by the two photons in separately deployed optical fibres.
The photon pairs are identified with a signal-to-noise ratio of at most {1.92} (Figure~\ref{fig:EPS_results}).
This constitutes a step towards an ultra-long fibre-based entanglement distribution, which so far has been performed over maximum distances of {248}~km in deployed fibre~\cite{NBB22} and {300}~km in a laboratory environment~\cite{IMT13}.
The investigated link has the potential to push these distance limits of entanglement QKD in a future, more mature implementation.
A number of measures can be taken to improve the pair detection rate in order to enable the measurement of the polarisation entanglement. 
This includes, but is not limited to, noise reduction, compensation for chromatic dispersion for improved temporal filtering, improved system alignment with a laser for optimised detection, and the utilisation of the full EPS spectrum for improved source brightness.

\section{Conclusions}

\noindent The performed experiments highlight the excellent potential of the euNetworks {224}~km submarine optical fibre to connect the United Kingdom and Ireland via quantum communications for the first time.
Our results show that protocols ranging from QKD to entanglement distribution, exploiting optical phase or polarisation for encoding, are feasible across the link if high-efficiency low-noise single-photon detectors are employed.
These preliminary experiments pave the way for others that are closer to a real-world implementation.
Given the industrial nature of the link used in the present set of experiments, this prospect is particularly desirable and is planned for the near future.  

\begin{acknowledgments}
\noindent We acknowledge invaluable help with the logistics from Chris Silk and Alex Van Der Arend, and the support received by the technical team of the Institute for Safe Autonomy.
Funding has been provided through the partnership resource scheme of the EPSRC Quantum Communications Hub grant (EP/T001011/1).
The authors declare no conflict of interest. 
The funders had no role in the design of the study; in the collection, analyses, or interpretation of data; in the writing of the manuscript; or in the decision to publish the~results.
\end{acknowledgments}


\begin{thebibliography}{0}%
\makeatletter
\providecommand \@ifxundefined [1]{%
 \@ifx{#1\undefined}
}%
\providecommand \@ifnum [1]{%
 \ifnum #1\expandafter \@firstoftwo
 \else \expandafter \@secondoftwo
 \fi
}%
\providecommand \@ifx [1]{%
 \ifx #1\expandafter \@firstoftwo
 \else \expandafter \@secondoftwo
 \fi
}%
\providecommand \natexlab [1]{#1}%
\providecommand \enquote  [1]{``#1''}%
\providecommand \bibnamefont  [1]{#1}%
\providecommand \bibfnamefont [1]{#1}%
\providecommand \citenamefont [1]{#1}%
\providecommand \href@noop [0]{\@secondoftwo}%
\providecommand \href [0]{\begingroup \@sanitize@url \@href}%
\providecommand \@href[1]{\@@startlink{#1}\@@href}%
\providecommand \@@href[1]{\endgroup#1\@@endlink}%
\providecommand \@sanitize@url [0]{\catcode `\\12\catcode `\$12\catcode `\&12\catcode `\#12\catcode `\^12\catcode `\_12\catcode `\%12\relax}%
\providecommand \@@startlink[1]{}%
\providecommand \@@endlink[0]{}%
\providecommand \url  [0]{\begingroup\@sanitize@url \@url }%
\providecommand \@url [1]{\endgroup\@href {#1}{\urlprefix }}%
\providecommand \urlprefix  [0]{URL }%
\providecommand \Eprint [0]{\href }%
\providecommand \doibase [0]{https://doi.org/}%
\providecommand \selectlanguage [0]{\@gobble}%
\providecommand \bibinfo  [0]{\@secondoftwo}%
\providecommand \bibfield  [0]{\@secondoftwo}%
\providecommand \translation [1]{[#1]}%
\providecommand \BibitemOpen [0]{}%
\providecommand \bibitemStop [0]{}%
\providecommand \bibitemNoStop [0]{.\EOS\space}%
\providecommand \EOS [0]{\spacefactor3000\relax}%
\providecommand \BibitemShut  [1]{\csname bibitem#1\endcsname}%
\let\auto@bib@innerbib\@empty
%</preamble>
\end{thebibliography}%


\begin{thebibliography}{999}

\bibitem[Bennett and Brassard()]{bennett_quantum_2014}
Bennett, C.H.; Brassard, G.
\newblock Quantum cryptography: Public key distribution and coin tossing.
\newblock {\em Theor. Comp. Sci.} \textbf{2014}, {\em 560},~7--11.
\newblock {\url{https://doi.org/10.1016/j.tcs.2014.05.025}}.

\bibitem[Gisin \em{et~al.}()Gisin, Ribordy, Tittel, and
  Zbinden]{gisin_quantum_2002}
Gisin, N.; Ribordy, G.; Tittel, W.; Zbinden, H.
\newblock Quantum cryptography.
\newblock {\em Rev. Mod. Phys.} \textbf{2002}, {\em 74},~145--{195.} 
\newblock 
  {\url{https://doi.org/10.1103/RevModPhys.74.145}}.

\bibitem[Xu \em{et~al.}()Xu, Ma, Zhang, Lo, and Pan]{xu_secure_2020}
Xu, F.; Ma, X.; Zhang, Q.; Lo, H.K.; Pan, J.W.
\newblock Secure quantum key distribution with realistic devices.
\newblock {\em Rev. Mod. Phys.} \textbf{2020}, {\em 92},~025002.
\newblock {\url{https://doi.org/10.1103/RevModPhys.92.025002}}.

\bibitem[Pirandola \em{et~al.}()Pirandola, Andersen, Banchi, Berta, Bunandar,
  Colbeck, Englund, Gehring, Lupo, Ottaviani, Pereira, Razavi, Shamsul~Shaari,
  Tomamichel, Usenko, Vallone, Villoresi, and Wallden]{pirandola_advances_2020}
Pirandola, S.; Andersen, U.L.; Banchi, L.; Berta, M.; Bunandar, D.; Colbeck,
  R.; Englund, D.; Gehring, T.; Lupo, C.; Ottaviani, C.;  et~al.
\newblock Advances in quantum cryptography.
\newblock {\em Adv. Opt. Photonics} \textbf{2020}, {\em 12},~1012.
\newblock {\url{https://doi.org/10.1364/AOP.361502}}.

\bibitem[Bennett \em{et~al.}(1992)Bennett, Bessette, Brassard, Salvail, and
  Smolin]{BCBB+92}
Bennett, C.H.; Bessette, F.; Brassard, G.; Salvail, L.; Smolin, J.
\newblock Experimental quantum cryptography.
\newblock {\em J. Cryptol.} {\bf 1992}, {\em 5},~3.
\newblock {\url{https://doi.org/10.1007/bf00191318}}.

\bibitem[Bennett(1992)]{B92}
Bennett, C.H.
\newblock Quantum cryptography using any two nonorthogonal states.
\newblock {\em Phys. Rev. Lett.} {\bf 1992}, {\em 68},~3121--3124.
\newblock {\url{https://doi.org/10.1103/PhysRevLett.68.3121}}.

\bibitem[Agnesi \em{et~al.}()Agnesi, Avesani, Stanco, Villoresi, and
  Vallone]{agnesi_all-fiber_2019}
Agnesi, C.; Avesani, M.; Stanco, A.; Villoresi, P.; Vallone, G.
\newblock All-fiber self-compensating polarization encoder for quantum key
  distribution.
\newblock {\em Opt. Lett.} \textbf{2019}, {\em 44},~2398--2401.
\newblock 
  {\url{https://doi.org/10.1364/OL.44.002398}}.

\bibitem[Avesani \em{et~al.}()Avesani, Agnesi, Stanco, Vallone, and
  Villoresi]{avesani_stable_2020}
Avesani, M.; Agnesi, C.; Stanco, A.; Vallone, G.; Villoresi, P.
\newblock Stable, low-error, and calibration-free polarization encoder for
  free-space quantum communication.
\newblock {\em Opt. Lett.} \textbf{2020}, {\em 45},~4706--4709.
\newblock 
  {\url{https://doi.org/10.1364/OL.396412}}.

\bibitem[Wang \em{et~al.}()Wang, Wang, Hu, Zapatero, Qian, Qi, Curty, and
  Lo]{wang_fully_2023}
Wang, W.; Wang, R.; Hu, C.; Zapatero, V.; Qian, L.; Qi, B.; Curty, M.; Lo, H.K.
\newblock Fully Passive Quantum Key Distribution.
\newblock {\em Phys. Rev. Lett.} \textbf{2023}, {\em 130},~220801.
\newblock 
  {\url{https://doi.org/10.1103/PhysRevLett.130.220801}}.

\bibitem[Lucamarini \em{et~al.}()Lucamarini, Yuan, Dynes, and
  Shields]{lucamarini_overcoming_2018}
Lucamarini, M.; Yuan, Z.L.; Dynes, J.F.; Shields, A.J.
\newblock Overcoming the rate-distance limit of quantum key distribution
  without quantum repeaters.
\newblock {\em Nature} \textbf{2018}, {\em 557},~400--403.
\newblock {\url{https://doi.org/10.1038/s41586-018-0066-6}}.

\bibitem[Minder \em{et~al.}()Minder, Pittaluga, Roberts, Lucamarini, Dynes,
  Yuan, and Shields]{minder_experimental_2019}
Minder, M.; Pittaluga, M.; Roberts, G.L.; Lucamarini, M.; Dynes, J.F.; Yuan,
  Z.L.; Shields, A.J.
\newblock Experimental quantum key distribution beyond the repeaterless secret
  key capacity.
\newblock {\em Nat. Photonics} \textbf{2019}, {\em 13},~334--338.
\newblock {\url{https://doi.org/10.1038/s41566-019-0377-7}}.

\bibitem[Pittaluga \em{et~al.}()Pittaluga, Minder, Lucamarini, Sanzaro,
  Woodward, Li, Yuan, and Shields]{pittaluga_600-km_2021}
Pittaluga, M.; Minder, M.; Lucamarini, M.; Sanzaro, M.; Woodward, R.I.; Li,
  M.J.; Yuan, Z.; Shields, A.J.
\newblock 600-km repeater-like quantum communications with dual-band
  stabilization.
\newblock {\em Nat. Photonics} \textbf{2021}, {\em 15},~530--535.
\newblock {\url{https://doi.org/10.1038/s41566-021-00811-0}}.

\bibitem[Wang \em{et~al.}()Wang, Yin, He, Chen, Wang, Ye, Zhou, Fan-Yuan, Wang,
  Chen, Zhu, Morozov, Divochiy, Zhou, Guo, and Han]{wang_twin-field_2022}
Wang, S.; Yin, Z.Q.; He, D.Y.; Chen, W.; Wang, R.Q.; Ye, P.; Zhou, Y.;
  Fan-Yuan, G.J.; Wang, F.X.; Chen, W.;  et~al.
\newblock Twin-field quantum key distribution over 830-km fibre.
\newblock {\em Nat. Photonics} \textbf{2022}, {\em 16},~154--161.
\newblock   {\url{https://doi.org/10.1038/s41566-021-00928-2}}.

\bibitem[Liu \em{et~al.}()Liu, Zhang, Jiang, Chen, Zhang, Pan, Ma, Dong, Xiong,
  Zhang, Li, Wang, Wu, Chen, You, Wang, Zhang, and Pan]{liu_experimental_2023}
Liu, Y.; Zhang, W.J.; Jiang, C.; Chen, J.P.; Zhang, C.; Pan, W.X.; Ma, D.;
  Dong, H.; Xiong, J.M.; Zhang, C.J.;  et~al.
\newblock Experimental Twin-Field Quantum Key Distribution over 1000 km Fiber
  Distance.
\newblock {\em Phys. Rev. Lett.} \textbf{2023}, {\em 130},~210801.
\newblock 
  {\url{https://doi.org/10.1103/PhysRevLett.130.210801}}.

\bibitem[Pirandola \em{et~al.}(2017)Pirandola, Laurenza, Ottaviani, and
  Banchi]{pirandola_fundamental_2017}
Pirandola, S.; Laurenza, R.; Ottaviani, C.; Banchi, L.
\newblock Fundamental Limits of Repeaterless Quantum Communications.
\newblock {\em Nat. Commun.} {\bf 2017}, {\em 8},~15043.
\newblock {\url{https://doi.org/10.1038/ncomms15043}}.

\bibitem[Ekert()]{ekert_quantum_1991}
Ekert, A.K.
\newblock Quantum cryptography based on Bell's theorem.
\newblock {\em Phys. Rev. Lett.} \textbf{1991}, {\em 67},~661--663.
\newblock 
  {\url{https://doi.org/10.1103/PhysRevLett.67.661}}.

\bibitem[Bennett \em{et~al.}({\natexlab{a}})Bennett, Brassard, and
  Mermin]{bennett_quantum_1992}
Bennett, C.H.; Brassard, G.; Mermin, N.D.
\newblock Quantum cryptography without Bell's theorem.
\newblock {\em Phys. Rev. Lett.} \textbf{1992}, {\em 68},~557--559.
\newblock 
  {\url{https://doi.org/10.1103/PhysRevLett.68.557}}.

\bibitem[Bennett \em{et~al.}({\natexlab{b}})Bennett, Brassard, Crépeau, Jozsa,
  Peres, and Wootters]{bennett_teleporting_1993}
Bennett, C.H.; Brassard, G.; Crépeau, C.; Jozsa, R.; Peres, A.; Wootters, W.K.
\newblock Teleporting an unknown quantum state via dual classical and
  Einstein--Podolsky-Rosen channels.
\newblock {\em Phys. Rev. Lett.} \textbf{1993}, {\em 70},~1895--1899.
\newblock 
  {\url{https://doi.org/10.1103/PhysRevLett.70.1895}}.

\bibitem[Bouwmeester \em{et~al.}()Bouwmeester, Pan, Mattle, Eibl, Weinfurter,
  and Zeilinger]{bouwmeester_experimental_1997}
Bouwmeester, D.; Pan, J.W.; Mattle, K.; Eibl, M.; Weinfurter, H.; Zeilinger, A.
\newblock Experimental quantum teleportation.
\newblock {\em Nature} \textbf{1997}, {\em 390},~575--579.
\newblock 
  {\url{https://doi.org/10.1038/37539}}.

\bibitem[Boschi \em{et~al.}()Boschi, Branca, De~Martini, Hardy, and
  Popescu]{boschi_experimental_1998}
Boschi, D.; Branca, S.; De~Martini, F.; Hardy, L.; Popescu, S.
\newblock Experimental Realization of Teleporting an Unknown Pure Quantum State
  via Dual Classical and Einstein--Podolsky-Rosen Channels.
\newblock {\em Phys. Rev. Lett.} \textbf{1998}, {\em 80},~1121--1125.
\newblock 
  {\url{https://doi.org/10.1103/PhysRevLett.80.1121}}.

\bibitem[Peev \em{et~al.}(2009)Peev, Pacher, All{\'{e}}aume, Barreiro, Bouda,
  Boxleitner, Debuisschert, Diamanti, Dianati, Dynes, Fasel, Fossier, FÃ¼rst,
  Gautier, Gay, Gisin, Grangier, Happe, Hasani, Hentschel, HÃ¼bel, Humer,
  LÃ¤nger, Legr{\'{e}}, Lieger, Lodewyck, LorÃ¼nser, LÃ¼tkenhaus,
  Marhold, Matyus, Maurhart, Monat, Nauerth, Page, Poppe, Querasser, Ribordy,
  Robyr, Salvail, Sharpe, Shields, Stucki, Suda, Tamas, Themel, Thew, Thoma,
  Treiber, Trinkler, Tualle-Brouri, Vannel, Walenta, Weier, Weinfurter,
  Wimberger, Yuan, Zbinden, and Zeilinger]{PPA+09}
Peev, M.; Pacher, C.; All{\'{e}}aume, R.; Barreiro, C.; Bouda, J.; Boxleitner,
  W.; Debuisschert, T.; Diamanti, E.; Dianati, M.; Dynes, J.F.;  et~al.
\newblock The {SECOQC} quantum key distribution network in Vienna.
\newblock {\em New J. Phys.} {\bf 2009}, {\em 11},~075001.
\newblock {\url{https://doi.org/10.1088/1367-2630/11/7/075001}}.

\bibitem[Sasaki \em{et~al.}(2011)Sasaki, Fujiwara, Ishizuka, Klaus, Wakui,
  Takeoka, Miki, Yamashita, Wang, Tanaka, Yoshino, Nambu, Takahashi, Tajima,
  Tomita, Domeki, Hasegawa, Sakai, Kobayashi, Asai, Shimizu, Tokura, Tsurumaru,
  Matsui, Honjo, Tamaki, Takesue, Tokura, Dynes, Dixon, Sharpe, Yuan, Shields,
  Uchikoga, Legr{\'{e}}, Robyr, Trinkler, Monat, Page, Ribordy, Poppe,
  Allacher, Maurhart, LÃ¤nger, Peev, and Zeilinger]{SFI+11}
Sasaki, M.; Fujiwara, M.; Ishizuka, H.; Klaus, W.; Wakui, K.; Takeoka, M.;
  Miki, S.; Yamashita, T.; Wang, Z.; Tanaka, A.;  et~al.
\newblock Field test of quantum key distribution in the Tokyo {QKD} Network.
\newblock {\em Opt. Express} {\bf 2011}, {\em 19},~10387.
\newblock {\url{https://doi.org/10.1364/oe.19.010387}}.

\bibitem[Dynes \em{et~al.}(2019)Dynes, Wonfor, Tam, Sharpe, Takahashi,
  Lucamarini, Plews, Yuan, Dixon, Cho, Tanizawa, Elbers, Greißer, White,
  Penty, and Shields]{DWT+19}
Dynes, J.F.; Wonfor, A.; Tam, W.W.S.; Sharpe, A.W.; Takahashi, R.; Lucamarini,
  M.; Plews, A.; Yuan, Z.L.; Dixon, A.R.; Cho, J.;  et~al.
\newblock Cambridge quantum network.
\newblock {\em NPJ Quantum Inf.} {\bf 2019}, {\em 5},~101.
\newblock {\url{https://doi.org/10.1038/s41534-019-0221-4}}.

\bibitem[Joshi \em{et~al.}(2020)Joshi, Aktas, Wengerowsky, Lončarić, Neumann,
  Liu, Scheidl, Lorenzo, Željko Samec, Kling, Qiu, Razavi, Stipčević,
  Rarity, and Ursin]{JAW+19}
Joshi, S.K.; Aktas, D.; Wengerowsky, S.; Lončarić, M.; Neumann, S.P.; Liu,
  B.; Scheidl, T.; Lorenzo, G.C.; Željko Samec.; Kling, L.;  et~al.
\newblock A trusted node–free eight-user metropolitan quantum communication
  network.
\newblock {\em Sci. Adv.} {\bf 2020}, {\em 6},~eaba0959. 
\newblock {\url{https://doi.org/10.1126/sciadv.aba0959}}.

\bibitem[Avesani \em{et~al.}(2021)Avesani, Calderaro, Foletto, Agnesi,
  Picciariello, Santagiustina, Scriminich, Stanco, Vedovato, Zahidy, Vallone,
  and Villoresi]{Avesani:21}
Avesani, M.; Calderaro, L.; Foletto, G.; Agnesi, C.; Picciariello, F.;
  Santagiustina, F.B.L.; Scriminich, A.; Stanco, A.; Vedovato, F.; Zahidy, M.;
  et~al.
\newblock Resource-effective quantum key distribution: A field trial in Padua
  city center.
\newblock {\em Opt. Lett.} {\bf 2021}, {\em 46},~2848--2851.
\newblock {\url{https://doi.org/10.1364/OL.422890}}.

\bibitem[Wonfor \em{et~al.}(2021)Wonfor, White, Lord, Nejabati, Spiller, Dynes,
  Shields, and Penty]{WWL+21}
Wonfor, A.; White, C.; Lord, A.; Nejabati, R.; Spiller, T.P.; Dynes, J.F.;
  Shields, A.J.; Penty, R.V.
\newblock Quantum networks in the {UK}.
\newblock In \emph{Metro and Data Center Optical Networks and  Short-Reach Links {IV}}; Glick, M., Srivastava, A.K., Akasaka, Y., Eds.;  {SPIE:} Bellingham, WA,  USA, 
  2021.
\newblock {\url{https://doi.org/10.1117/12.2578598}}.

\bibitem[Ribezzo \em{et~al.}(2023)Ribezzo, Zahidy, Vagniluca, Biagi,
  Francesconi, Occhipinti, Oxenl{\o}we, Lon{\v c}ari{\'c}, Cviti{\'c}, Stip{\v
  c}evi{\'c}, Pu{\v s}avec, Kaltenbaek, Ram{\v s}ak, Cesa, Giorgetti, Scazza,
  Bassi, De~Natale, Cataliotti, Inguscio, Bacco, and
  Zavatta]{ribezzo_deploying_2023}
Ribezzo, D.; Zahidy, M.; Vagniluca, I.; Biagi, N.; Francesconi, S.; Occhipinti,
  T.; Oxenl{\o}we, L.K.; Lon{\v c}ari{\'c}, M.; Cviti{\'c}, I.; Stip{\v
  c}evi{\'c}, M.;  et~al.
\newblock Deploying an {{Inter-European Quantum Network}}.
\newblock {\em Adv. Quantum Technol.} {\bf 2023}, {\em 6},~2200061.
\newblock {\url{https://doi.org/10.1002/qute.202200061}}.

\bibitem[Bersin \em{et~al.}(2023)Bersin, Grein, Sutula, Murphy, Huan, Stevens,
  Suleymanzade, Lee, Riedinger, Starling, Stas, Knaut, Sinclair, Assumpcao,
  Wei, Knall, Machielse, Sukachev, Levonian, Bhaskar, Lon{\v c}ar, Hamilton,
  Lukin, Englund, and Dixon]{bersin_development_2023}
Bersin, E.; Grein, M.; Sutula, M.; Murphy, R.; Huan, Y.Q.; Stevens, M.;
  Suleymanzade, A.; Lee, C.; Riedinger, R.; Starling, D.J.;  et~al.
\newblock Development of a {{Boston-area}} 50-Km Fiber Quantum Network Testbed.
   \emph{arXiv} \textbf{2023},  arXiv:2307.15696.

\bibitem[Neumann \em{et~al.}()Neumann, Buchner, Bulla, Bohmann, and
  Ursin]{NBB22}
Neumann, S.P.; Buchner, A.; Bulla, L.; Bohmann, M.; Ursin, R.
\newblock Continuous entanglement distribution over a transnational 248-km
  fiber link.
\newblock {\em Nat. Commun.} \textbf{2022}, {\em 13},~6134.
\newblock {\url{https://doi.org/10.1038/s41467-022-33919-0}}.

\bibitem[Wengerowsky \em{et~al.}({\natexlab{a}})Wengerowsky, Joshi,
  Steinlechner, Zichi, Dobrovolskiy, van~der Molen, Los, Zwiller, Versteegh,
  Mura, Calonico, Inguscio, Hübel, Bo, Scheidl, Zeilinger, Xuereb, and
  Ursin]{wengerowsky_entanglement_2019}
Wengerowsky, S.; Joshi, S.K.; Steinlechner, F.; Zichi, J.R.; Dobrovolskiy,
  S.M.; van~der Molen, R.; Los, J.W.N.; Zwiller, V.; Versteegh, M.A.M.; Mura,
  A.;  et~al.
\newblock Entanglement distribution over a 96-km-long submarine optical fiber.
\newblock {\em Proc. Natl. Acad. Sci. USA} \textbf{2019}, {\em
  116},~6684--6688.
\newblock 
  {\url{https://doi.org/10.1073/pnas.1818752116}}.

\bibitem[Wengerowsky \em{et~al.}({\natexlab{b}})Wengerowsky, Joshi,
  Steinlechner, Zichi, Liu, Scheidl, Dobrovolskiy, Molen, Los, Zwiller,
  Versteegh, Mura, Calonico, Inguscio, Zeilinger, Xuereb, and
  Ursin]{wengerowsky_passively_2020}
Wengerowsky, S.; Joshi, S.K.; Steinlechner, F.; Zichi, J.R.; Liu, B.; Scheidl,
  T.; Dobrovolskiy, S.M.; Molen, R.v.d.; Los, J.W.N.; Zwiller, V.;  et~al.
\newblock Passively stable distribution of polarisation entanglement over 192
  km of deployed optical fibre.
\newblock {\em NPJ Quantum Inf.} \textbf{2020}, {\em 6},~1--5.
\newblock 
  {\url{https://doi.org/10.1038/s41534-019-0238-8}}.

\bibitem[Ribezzo \em{et~al.}(2023)Ribezzo, Zahidy, Lemmi, Petitjean,
  De~Lazzari, Vagniluca, Conca, Tosi, Occhipinti, Oxenl{\o}we, Xuereb, Bacco,
  and Zavatta]{ribezzo_quantum_2023}
Ribezzo, D.; Zahidy, M.; Lemmi, G.; Petitjean, A.; De~Lazzari, C.; Vagniluca,
  I.; Conca, E.; Tosi, A.; Occhipinti, T.; Oxenl{\o}we, L.K.;  et~al.
\newblock Quantum {{Key Distribution}} over 100 Km Underwater Optical Fiber
  Assisted by a {{Fast-Gated Single-Photon Detector}}. \emph{arXiv}   \textbf{2023}, arXiv:2303.01449.

\bibitem[EUN()]{EUN}
euNetworks.
\newblock Available online: \url{https://eunetworks.com/} (accessed on 25 September 2023).


\bibitem[Clivati \em{et~al.}(2020)Clivati, Marra, Levi, Mura, Xuereb, and
  Calonico]{clivati_optical_2020}
Clivati, C.; Marra, G.; Levi, F.; Mura, A.; Xuereb, A.; Calonico, D.
\newblock Optical Frequency Transfer over Submarine Fibers.
\newblock In \emph{Proceedings of the Conference on Lasers and Electro-Optics (2020)};  Paper {SM}2N.1;  {Optica Publishing Group:} Washington, DC,  USA, 
  2020; p. SM2N.1.
\newblock {\url{https://doi.org/10.1364/CLEO_SI.2020.SM2N.1}}.

\bibitem[ozo()]{ozoptics_polarization_2023}
Polarization Entangled Photon Sources{\textbar}{OZ} Optics Ltd.
\newblock
  Available online: \url{https://www.ozoptics.com/products/polarization-entangled-photon-sources.html} (accessed on 25 September 2023).

\bibitem[Fr{\"o}hlich \em{et~al.}(2017)Fr{\"o}hlich, Lucamarini, Dynes,
  Comandar, Tam, Plews, Sharpe, Yuan, and Shields]{FLD+17}
Fr{\"o}hlich, B.; Lucamarini, M.; Dynes, J.F.; Comandar, L.C.; Tam, W.W.S.;
  Plews, A.; Sharpe, A.W.; Yuan, Z.; Shields, A.J.
\newblock Long-distance quantum key distribution secure against coherent
  attacks.
\newblock {\em Optica} {\bf 2017}, {\em 4},~163, 
\newblock {\url{https://doi.org/10.1364/optica.4.000163}}.

\bibitem[IDQ()]{IDQ_SNSPD}
ID281 Superconducting Nanowire Series{\textbar}IDQuantique.
\newblock
  Available online: \url{https://www.idquantique.com/quantum-sensing/products/id281-snspd-series/} (accessed on 25 September 2023).

\bibitem[Hwang(2003)]{Hwa03}
Hwang, W.Y.
\newblock Quantum key distribution with high loss: toward global secure
  communication.
\newblock {\em Phys. Rev. Lett.} {\bf 2003}, {\em 91},~057901.
\newblock
  {\url{https://doi.org/https://doi.org/10.1103/PhysRevLett.91.057901}}.

\bibitem[Wang(2005)]{Wan05}
Wang, X.B.
\newblock Beating the Photon-Number-Splitting Attack in Practical Quantum
  Cryptography.
\newblock {\em Phys. Rev. Lett.} {\bf 2005}, {\em 94},~230503.
\newblock {\url{https://doi.org/10.1103/physrevlett.94.230503}}.

\bibitem[Lo \em{et~al.}(2005)Lo, Ma, and Chen]{LMC05}
Lo, H.K.; Ma, X.; Chen, K.
\newblock Decoy State Quantum Key Distribution.
\newblock {\em Phys. Rev. Lett.} {\bf 2005}, {\em 94},~230504.
\newblock {\url{https://doi.org/10.1103/physrevlett.94.230504}}.

\bibitem[Horn and Jennewein(2019)]{HJ19}
Horn, R.; Jennewein, T.
\newblock Auto-balancing and robust interferometer designs for polarization
  entangled photon sources.
\newblock {\em Opt. Express} {\bf 2019}, {\em 27},~17369.
\newblock {\url{https://doi.org/10.1364/oe.27.017369}}.

\bibitem[Li \em{et~al.}()Li, Zhang, Lu, Li, Jiang, Liu, Huang, Li, Wang, Wang,
  Zhang, You, Xu, and Pan]{li_twin-field_2023}
Li, W.; Zhang, L.; Lu, Y.; Li, Z.P.; Jiang, C.; Liu, Y.; Huang, J.; Li, H.;
  Wang, Z.; Wang, X.B.;  et~al.
\newblock Twin-Field Quantum Key Distribution without Phase Locking.
\newblock {\em Phys. Rev. Lett.} \textbf{2023}, {\em 130},~250802.
\newblock 
  {\url{https://doi.org/10.1103/PhysRevLett.130.250802}}.

\bibitem[Clivati \em{et~al.}()Clivati, Meda, Donadello, Virzì, Genovese, Levi,
  Mura, Pittaluga, Yuan, Shields, Lucamarini, Degiovanni, and
  Calonico]{clivati_coherent_2022}
Clivati, C.; Meda, A.; Donadello, S.; Virzì, S.; Genovese, M.; Levi, F.; Mura,
  A.; Pittaluga, M.; Yuan, Z.; Shields, A.J.;  et~al.
\newblock Coherent phase transfer for real-world twin-field quantum key
  distribution.
\newblock {\em Nat. Commun.} \textbf{2022}, {\em 13},~157.
\newblock 
  {\url{https://doi.org/10.1038/s41467-021-27808-1}}.

\bibitem[Ma \em{et~al.}()Ma, Qi, Zhao, and Lo]{ma_practical_2005}
Ma, X.; Qi, B.; Zhao, Y.; Lo, H.K.
\newblock Practical decoy state for quantum key distribution.
\newblock {\em Phys. Rev. A} \textbf{2005}, {\em 72},~012326.
\newblock {\url{https://doi.org/10.1103/PhysRevA.72.012326}}.

\bibitem[Inagaki \em{et~al.}(2013)Inagaki, Matsuda, Tadanaga, Asobe, and
  Takesue]{IMT13}
Inagaki, T.; Matsuda, N.; Tadanaga, O.; Asobe, M.; Takesue, H.
\newblock Entanglement distribution over 300 km of fiber.
\newblock {\em Opt. Express} {\bf 2013}, {\em 21},~23241--23249.
\newblock {\url{https://doi.org/10.1364/OE.21.023241}}.

\end{thebibliography}
\end{document}